# Reviewers of Educational Immersive and Extended Reality (XR) experiences: Who is creating these reviews and why?

Dr Sophie Mckenzie**
Senior Lecturer
sophie.mckenzie@deakin.edu.au
Phone: +61352271263
School of Information Technology, Deakin University, Geelong, Australia

Dr Shaun Bangay
Senior Lecturer
shaun.bangay@deakin.edu.au
School of Information Technology, Deakin University, Geelong, Australia

Dr Maria Nicholas
Senior Lecturer
maria.nicholas@deakin.edu.au
School of Education, Deakin University, Geelong, Australia;

Dr Adam Cardilini
Lecturer
adam.cardilini@deakin.edu.au
School of Life and Environmental Sciences, Deakin University, Geelong, Australia

Mr Manjeet Singh
PhD Candidate
guy.woodbradley@deakin.edu.au
School of Information Technology, Deakin University, Geelong, Australia

*corresponding author

## 1. Introduction

This paper presents a scoping review of literature to examine who is reviewing educational immersive or extended reality (eduXR) experiences and why. EduXR experiences in augmented, virtual or mixed reality take many forms, from supporting manual training, engaging learners in conservation, to provide opportunities for social connection (Zheng 2018, Bangay et al.,). For users of eduXR, reviews of an experience can provide information that helps them determine whether it will meet their learning needs or not. The 'source' of the review, that is, who they are and why they have conducted the review, is critical in helping the user judge the review's quality and relevance. At present, there is no settled review system in place for eduXR, though relevant frameworks exist for serious games review with relevance and overlap for some, but not all, eduXR experiences. While some authors have engaged in preparing a detailed review structure for eduXR (Santos et al., 2019), there remains a need for a clear and simple way for users of eduXR to know details about reviewers, e.g., who and why, to help make it easier for users to identify relevant reviews and gain useful insight about eduXR experiences. To help address this issue, we conducted a scoping review asking the question; **Who is creating eduXR reviews, and why?** We identified 16 papers that present an academic evaluation on the review process of educational immersive or extended reality (eduXR) reviews. The 16 papers were analysed, coding for 'who' themes and 'why' themes over two separate cycles, using thematic analysis. An analysis looked to examine what we know regarding who is providing the reviews, and why, to help us to understand what enables, inhibits and what is yet unknown about how the eduXR community goes about making informed choices regarding the eduXR experiences they engage with.

## 2. Literature: What is an EduXR Review?

Educational immersive or extended reality (eduXR) experiences are presented in augmented, virtual or mixed reality with a focus on providing an educational experience. Although there are similarities, eduXR experience differ from games or serious games in their presentation style, with varying degrees of synthetic content provided to the user on a spectrum from overlaying the real world to fully immersive (in a head mounted display). In terms of content, eduXR and serious games have synergies, particularly in terms of having a goal to engage with learners and a defined learning objective (Bedwell et al., 2012). Yet eduXR and serious games can be considered distinct based on their interaction mechanisms, affordances and metaphors, which can diverge in relation to game elements employed. Another important aspect of eduXR is that the user base tend to have different motivations and characteristics to those playing games for entertainment. In particular, eduXR users will be looking for experiences that can help them support or achieve learning outcomes. Normally, game reviews provide users the opportunity to determine if they would enjoy playing a game. For eduXR users reviews need to relay details about the learning experience to help people determine if it is an appropriate learning tool. Additionally, this must be done for people who may be unexperienced with XR or evaluating games. This makes eduXR reviews and their content particularly important for users to make informed decisions about eduXR experiences.

There are examples of how a 'review' of an eduXR experience can be used to inform both the entertainment and educational quality of an immersive experience (Eberhard et al., 2018; Kasper et al., 2019; Lin et al., 2019; Wang & Goh, 2020). In the area of serious games, numerous studies have considered how to suitably evaluate the user experience, with various characteristics being identified that should be included in a review and that categorise reviews (Coleman & Talib, 2014; Ho & Tu, 2012; Kirschner & Williams, 2014). The reviews of eduXR experiences are yet to receive the same level of scrutiny despite having educational qualities that make the reviewer and their motivations an important aspect of the review process, e.g., the difference between a teacher and a student reviewing the same game matters.

While EduXR reviews have not received much attention research has investigated and described the characteristics of game reviews, the insights they provide about game experience, and their helpfulness to readers. Lin et al. (2019) analysed the general characteristics of reviews for over

6,000 video games available on the steam platform. On average reviews were about 200 words but were often longer for reviews of paid games. Reviews talked about the pros and cons of the game, made suggestions on improvements, or highlighted bugs. Players spent 13.5 hours on average playing a paid game before posting a review, with negative reviews often provided after less time playing a game. Players only spent on average one hour playing a free game before making a review.

Wang & Goh (2020) used online game reviews to collect feedback on game experience. They note that online reviews are voluminous, but often unstructured which can complicate their use as a formal feedback tool. Various game experience frameworks are included in their study, demonstrating 57 distinct components in which we can evaluate a video game. To analyse over 5000 online reviews (Wang & Goh, 2020) used topic modelling and sentiment analysis to find seven topics were mentioned throughout the reviews, these include: achievement, narrative, social interaction, social influence, visual/value, accessories, and general experience. (Koehler et al., 2017) used (Bedwell et al., 2012) taxonomy of game elements to categorise online gaming reviews. The taxonomy explores the cognitive, motivational, and affective outcomes of gaming. It has nine overarching dimensions: Action Language, Assessment, Conflict/Challenge, Control, Environment, Game Fiction, Human Interaction, Immersion, and Rules/Goals. Through a pilot review process (Koehler et al., 2017) added the dimensions of value judgement and comparison to the list. (Koehler et al., 2017) assessed 200 reviews on VideoGameGeek using the 11 dimensions. In these reviews Rules/Goals, Immersion, Conflict/Challenge, and Game Fiction appeared most frequently, conversely, Human Interaction, Assessment, Environment, and Control were mentioned least in the reviews. Value-judgments and comparison also rated highly in the reviews. When a game was rated highly it mentioned Game Fiction, in contrast, reviews of low-rated games had increased commentary about Immersion and Conflict/Challenge. In 2012, (Ho & Tu, 2012) analysed the important characteristics of mobile games and found that fun, information-richness, perceived value, after sales services, stableness, and challenge were mentioned the most out of 1485 reviews.

Eberhard et al. (2018) analysed the usefulness of game reviews on the steam platform. From over 100,000 reviews, they found that review length and gameplay time were significant factors that contributed to the helpfulness of the review, as determined by steam user votes on reviews. They also found that not all reviews tagged on the steam platform were necessarily useful, and that outside factors (such as the authors status) may have played a part. Wang & Goh (2020) found that shorter reviews tended to be associated with a positive attribute towards the game, whereas longer reviews tended to be negative. Conversely, (Livingston et al., 2011) explored the impact of games reviews on a player's experience of a game. They found that the nature of the game review, either positive or negative, did not have a direct impact on the game experience for the player. Rather review valence is a post-play cognitive rationalization of the experience with the content of the review. Kasper et al. (2019) reviewed genre, score and review text as related to the helpfulness of 319,017 video game reviews on Metacritic. Broadly they found that across genres that review score impacts the user perception on the helpfulness of the review, however this can vary across genres. In some instances, review score is not significant in the user perception of usefulness but rather the review text helps validate review use.

The above shows that given the unstructured and inconsistent format of game reviews, reviewer time spent playing the game and the concise nature of reviews played a role in helping users determine the usefulness of a review. Also, game reviews can provide insight into key topics that help to describe games and understand user experience. For eduXR we would expect additional important topics (e.g., quality of learning outcomes, level of learning competency) would need to be covered for the reviews to be helpful.

*2.1 How are eduXR reviews reported?*

eduXR as a relatively new and emerging platform, resulting in few analyses of eduXR reviews in the literature. However, there has been work published which attempts to describe formal evaluations of eduXR experiences. We believe these more academic evaluations can provide insight into review processes. Combined with work done on serious games, of which eduXR is arguable a subcategory (though some eduXR would fall outside of this, e.g., training simulations), we draw upon these papers to inform our understanding of eduXR reviews.

When reviewing serious games, Calderón & Ruiz (2015) argued that you can organize an review into two stages: 1) the type and application domain of serious games and 2) the review method and associated characteristics. Using these stages, Calderón & Ruiz (2015) found that over 60% of serious games in their study were evaluated for use in the higher education context, with the remaining 40% evaluated in either a primary or secondary context. When looking at the sample size of the interventions evaluated, on average the sample contained around 40 learners when evaluating an experience. The domains of serious games evaluated by Calderón & Ruiz (2015) include; health and wellbeing, cultural training, professional learning and training, social skills training, support and assist life decisions, and education/ formal knowledge across various areas. Of these domains, only a handful of papers were available that report on using serious games via a virtual world to provide the experience.

Critically, when games and experiences are evaluated in an academic setting often self-report questionnaires are the main method in which a user or player will provide their evaluation. For example, when evaluating serious games, Calderón & Ruiz (2015) found the following evaluation methods in use; questionnaires, interviews, logs, discussions, videos, frameworks, observations, or other methods. Yáñez-Gómez et al. (2017) conducted a systematic literature review of 187 papers that considered usability evaluation of serious games. They found that evaluations were largely conducted using ad-hoc questionnaires with general users, where the user completed exploratory actions to investigate the game. Task driven evaluations using formal techniques and conducted with experts did not frequent the literature. Neilson's Heuristic checklist was the most frequently used evaluation tool. Faizan et al. (2019) conducted a literature review to analyse evaluation methods for the pre-game, in-game, and post-game assessment of simulation games. From their literature they identified 37 different assessment types used pre-game, in-game and post-game, with questionnaires remaining a dominant approach to gathering user feedback from a game experience.

Other studies have explored how serious games are reviewed. Calderón & Ruiz (2015) noted the following characteristics were used; the aesthetics of the games design, the user's satisfaction, the usability, usefulness, understandability, the impact on motivation, performance, playability, pedagogical aspects, learning outcomes, engagement, user's experience, efficacy on outcomes, social impact, cognitive behavior, enjoyment. Yáñez-Gómez et al. (2017) defined usability evaluations of serious games to consist measure to assess satisfaction, efficiency and effectiveness. Coleman & Talib (2014) describe in their serious game review criteria the importance of context and elements when describing a game. Context includes type of game, genre, the type of virtual environment, gameplay synopsis, learning purpose, target users, players expected motivation, how the game is intended to be used and intended outcomes. Elements include a discussion of the learning objectives and game goals, narrative, distinguishing game characteristics, instructional methods, game mechanics, immersive elements, assessment and feedback. (Coleman & Talib, 2014) also include a design analysis as a part of a review, using five questions to prompt review response:

1. How well does the game design achieve the intended instructional purposes?
2. How well does the game balance instruction and gameplay?
3. How appropriate is the level of authenticity?
4. How well does the game maintain learner motivation?
5. How well does the game appeal to its intended role in the learning environment?

Petri & Gresse von Wangenheim (2017) completed a systematic literature review on how to review educational games. They analysed how the review approaches are defined (quality factors, theoretical constructs), operationalized (research designs, data collection instruments, data analysis methods), how they have been developed (development methodology) and evaluated (evaluated aspects, number of applications & data points and data analysis methods). When considering learning as a quality factor, Petri & Gresse von Wangenheim (2017) found that competence and learner performance were the current methods used for evaluating any learning outcomes. Overall, they argue that a review should include the pedagogical relevant information, including context, environment, learner specifications, preferences, game play, and user experience. Caserman et al. (2020) also describes the quality criteria for serious games, focusing on the attractive and effective elements of games. They argue that in a review criterion characterizing the game goal is very important, along with domain appropriate content, appropriate feedback on progress and rewards. Along with goals, a game review should also focus on game enjoyment which includes engagement, flow control, emotional connection, social interaction, need for immersive experience, and appropriate sounds and graphics. The technology for which the game was developed should be deemed suitable by the target audience, with intuitive game mechanics and natural mapping also required. The quality of the game is also verified through proof of effectiveness and sustainable effects along with awards and rating given to the game.

*2.2 The Audience of eduXR reviews*

It is important to consider how the audience may use a review, and their agency over the information in the review. Hendrickx (2022) defines audience agency as either deliberate or incidental. Within these two types of audience agency occuring on three levels: the micro level of the individual, the meso level of organised groups and the macro level of collective audiences. Of particularly interest to this discussion is deliberative and incidental, individual and micro level agency, which describes individuals exerting action within existing media structures and socio-political climates (Hendrickx, 2022). Individual and deliberate micro agency is shown in things like "sharing a news article, commenting on a post, or signing an online petition all signify deliberate actions of agency for various types of media content" (Hendrickx, 2022, p. 4). Incidental agency may include liking or up-voting certain posts, with the nature of this moving to deliberate agency depending on the audience's position. In this study, our focus on academic evaluations of eduXR enables us to explore what is currently available for the audience, and what is lacking in reviews, to enable better understandings of how to support individual (micro) audience agency and engagement more effectively.

At present, the education community is lacking a formal process for reviewing educational immersive or extended reality (eduXR) experiences in ways that will help readers determine the educational suitability of the experience. To assist in exploring this issue, we conducted a scoping review to provide some insight into what is currently known about the purpose/s and use of eduXR and gaming reviews as determined in academic literature – specifically educational immersive or extended reality (eduXR) gaming reviews. In particular, we explored why authors may be creating such reviews. The objective of engaging in this scoping review is to initiate a conversation into the possible formalization of the eduXR review process for the education community.

## 3. Method

We conducted a scoping review (Tricco et al. 2018, p. 467), a method used to examine the "range (variety), and nature (characteristics) of the evidence of a topic" that will assist in planning for future research. Our scoping review protocol was similar to (Delgado & Der Ananian, 2021). The method used for this scoping review largely followed the process outlined in the PRISMA-ScR (Preferred Reporting Items for Systematic reviews and Meta-Analyses extension for Scoping Reviews) Checklist and Explanation (Tricco et al., 2018) (p.471), as shown in Table 1.

Table 1: Method Overview

|   | Task | PRISMA-ScR Task Description |
|---|---|---|
| 1 | Select information sources | Identify the databases you will use and dates of coverage |
| 2 | Establish search terms | Identify the search terms and limits you will use |
| 3 | Establish selection and exclusion process | Determine a process for selecting sources to include in the scoping review, including inclusion and exclusion criteria. |
| 4 | Establish data charting process | Establish a method for charting the data |
| 5 | Synthesise results | Establish a method for summarising the charted data. |

For this scoping review we searched the Web of Science database and Scopus using the search terms "game OR gaming AND review*" and "gaming review" from 2000-June 2022. We chose the search terms of 'game/ gaming review' as these terms are sufficiently broad and will include any eduXR review papers currently available (as eduXR can be considered a subcategory of serious games). Using Web of Science the search results were screened in groups of 50, with a total of 350 results reviewed. The search was halted when a page of 50 entries yielded zero relevant results (n = 350 entries), which was found to be page 7 of the search results. An entry was deemed relevant if its title or abstract reviewed or evaluated a digital game/s or investigated the features of a gaming review. The outcomes of this search are shown in table 2, with the page and entry number for each paper listed. To verify the outcome, Scopus was then searched using the same search string and date range. The first 200 results yielded only two relevant entries, and these entries also appeared in the Web of Science search. Only one relevant paper from the Scopus search was added to the entries (table 2). This data was inputted into separated Excel spreadsheets and NVivo software used to define themes and subthemes (nodes and sub-nodes) giving valuable insights to the qualitative data created by all the authors independently and cross referenced until agreement was met. The abstract was cross referenced when additional information was needed. Literature reviews about gaming and/or associated disorders (e.g. game addiction, sleep disorders or gambling), or game developers, were excluded from the final selection. Table 2 shows the order in which the 16 relevant entries appeared.

Table 2: Papers sourced with Scoping Review Method

| **Page 1** | (2) Koehler, MJ; Arnold, B; Boltz, LO Jun, 2017, A Taxonomy Approach to Studying How Gamers Review Games, *Simulation and Gaming*, VOL 48 (3), pp.363-380 https://doi.org/10.1177/1046878117703680 | (12) Lin, DY; Bezemer, CP; (...); Hassan, AE Feb 2019, An empirical study of game reviews on the Steam platform, *Empirical Software Engineering*, VOL 24 (1), pp.170-207 https://doi.org/10.1007/s10664-018-9627-4 | (29) Wang, XH and Goh, DHL Mar, 2020, Components of game experience: An automatic text analysis of online reviews, *Entertainment Computing*, VOL 33 https://doi.org/10.1016/j.entcom.2019.100338 | (45) Petri, G and von Wangenheim, CG, 2016, How to Evaluate Educational Games: a Systematic Literature Review *Journal of Universal Computer Science*, VOL 22 (7), pp.992-1021 10.3217/jucs-022-07-0992 |
|---|---|---|---|---|
| **Page 2** | (53) Kirschner, D and Williams, JP, 2014, Measuring Video Game Reviews, Engagement Through Gameplay, *Simulation and Gaming*, VOL 45 (4-5), pp. 593-610 https://doi.org/10.1177/1046878114554185 | (72) SC and Tu, YC, 2012, The Investigation of Online Reviews of Mobile Games, *10th Workshop on E-Business*, VOL 108, pp.130-139 https://doi.org/10.1007/978-3-642-29873-8_13 | (75) Livingston, IJ; Nacke, LE and Mandryk, RL, 2011, Influencing Experience: The Effects of Reading Game Reviews on Player Experience, *10th International Conference on Entertainment Computing*, pp.89 https://doi.org/10.1007/978-3-642-24500-8_10 | |
| **Page 3** | (111) Yanez-Gomez, R; Cascado-Caballero, D and Sevillano, JL, 2017, Academic methods for usability evaluation of serious games: a systematic review, *Multimedia Tools and Applications*, VOL 76 (4), pp.5755-5784 https://doi.org/10.1007/s11042-016-3845-9 | (114) Calderon, A and Ruiz, M, 2015, A systematic literature review on serious games evaluation: An application to software project management, *Computers and Education*, VOL 87, pp.396-422 https://doi.org/10.1016/j.compedu.2015.07.011 | (126) Faizan, Nilüfer, et al. "Classification of evaluation methods for the effective assessment of simulation games: Results from a literature review." (2019): 19-33."Classification of evaluation methods for the effective assessment of simulation games: Results from a literature review." | (150) P; Koncar, P; (...); Gutl, C, 2019, On the Role of Score, Genre and Text in Helpfulness of Video Game Reviews on Metacritic Kasper, *6th International Conference on Social Networks Analysis, Management and Security (SNAMS)*, pp.75-82 10.1109/SNAMS.2019.8931866 |

|  |  |  | (2019): 19-33. 10.3991/ijep.v9i1.9948 |  |
|---|---|---|---|---|
| **Page 4** | (164) Shiratuddin, MF and Thabet, W, 2011, Utilising a 3D Game Engine to Develop a Virtual Design Review System, *Journal of Information Technology in Construction* VOL 16, pp.39-68 | (165) Eberhard, L; Kasper, P; (…); Gutl, C, 2018, *Investigating Helpfulness of Video Game Reviews on the Steam Platform,* Fifth International Conference on Social Networks Analysis, Management and Security, pp.43-50 10.1109/SNAMS.2018.8554542 |  |  |
| **Page 5** | (237) Petri, G and von Wangenheim, CG, 2017, How games for computing education are evaluated? A systematic literature review, *Computers and Education,* VOL 107, pp.68-90 https://doi.org/10.1016/j.compedu.2017.01.00 |  |  |  |
| **Page 6** | (274) Coleman, SL and Hussain, TS, 2015, Game Review Criteria, *Chapter published in Design and Development of Training Games: Practical Guidelines from a Multidisciplinary Perspective*, pp.337-346 |  |  |  |
| **Scopus** | (19) Caserman P, Hoffmann K and Gobel S, 2020, Quality Criteria for Serious Games: Serious Part, Game Part, and Balance, *JMIR Serious Games 10.2196/19037* |  |  |  |

## 4. Results

When analysing the 16 papers identified, the research question 'who is creating eduXR reviews, and why' initiated the thematic analysis process. The 16 papers were analysed, coding for 'who' themes and 'why' themes over two separate cycles, using thematic analysis (Creswell & Poth, 2017) where the articles' content was categorized according to words and phrases, as shown in Table 3. The authors analysed the data separately using an Excel spreadsheet and NVivo software to analyse according to the themes and subthemes, and then compared their coding during collaborative meetings, reaching agreement on any that differed in the first wave of coding.

Table 3: Themes identified in the data.

| Category | Theme | Subtheme |
|---|---|---|
| Who | Author | Academic |
| | | Everyday game user |
| | | End user |
| | | Practitioner/trainer |
| | Expertise | Not provided |
| | | Everyday gamer |
| | | Professional gamer |
| | | Scholarly academic |
| | | Impartial educator/trainer |
| | | Quantified (e.g. number of posts/games/hours played) |
| | Audience | Consumers/gamers |
| | | Game developers |
| | | Academics/researchers |
| | | Educators/practitioners/instructors |
| Why | Share opinions | Influence game play experience |
| | | Persuade to purchase |
| | | Give suggestions to game developers |
| | | Improve gamers communications |
| | Quality control | Improve game interface and interactivity |
| | | Enhance play experience |
| | Education aspects | Educational goals achievement |
| | | Knowledge acquisition |
| | | Motivation and engagement |
| | | Achieve learning outcomes |
| | Rate and measure | Insights to game play |
| | | Gaming principles |
| | | Commercial viability |
| | Game play reasoning | Gaming properties / strategies |
| | | Empirical and quantitative data |

Based on the collaborative coding outcomes, two main themes were derived from the scoping review outcomes.

*4.1 Who is creating EduXR/ Game reviews?*

Following our scoping review, we found that the main authors of reviews or evaluations of games, were academics undertaking an empirical study or everyday game users. Our scoping approach identified six articles that cited everyday gamers, customers or users as the author of the review or evaluation (Eberhard et al., 2018; Ho & Tu, 2012; Kasper et al., 2019; Koehler et al., 2017; Lin et al., 2019; Wang & Goh, 2020). Eight articles (two by the same authors) cited academics as the author/s of the review or evaluation, having provided an analysis of the information gathered from their participants (the "end users") who took part in their empirical studies (Calderón & Ruiz, 2015; Faizan et al., 2019; Kirschner & Williams, 2014; Livingston et al., 2011; Petri et al., 2016; Yáñez-Gómez et al., 2017) or as a result of having conducted a literature review and "workshops with domain experts" (Caserman et al., 2020). One article cited practitioners and people who provide training using the game as the author/s of the review or evaluation (Coleman & Talib, 2014).

Six articles provided details about the level of expertise of reviewers/evaluators. Koehler et al. (2017) stated that their reviewers were 'everyday gamers' as distinct from professional reviews or professional staff. Lin et al. (2019) used the similar term 'players', with expertise determined by displaying "The number of playing hours of the reviewed game, the number of played games, and the number of previously posted reviews by the reviewer" (p.4). Likewise, review expertise in Eberhard et al. (2018) was defined by the "number of products owned" by the reviewer and "time spent playing the respective game" (pp. 43; 45). In four instances, the reviewer was an academic/researcher (Calderón & Ruiz, 2015; Faizan et al., 2019; Livingston et al., 2011; Petri et al., 2016). In one instance (Coleman & Talib, 2014) the reviewer was an "impartial" educator or educators that "provided training using the game" (p.338).

Five studies simply referred to the reviewer as an 'end user' (Petri et al., 2016), 'user' (Kasper et al., 2019; Yáñez-Gómez et al., 2017) 'customer' or 'consumer' (Ho & Tu, 2012) (pp: 133; 131). Kasper et al., p. (2019, p. 77) raise the issue that "there is no verification if a user actually owns a game or has played it before writing the review", emphasizing that the level of expertise of such 'users' can at times, and on some platforms, remain unknown. While Yáñez-Gómez et al. (2017) attest to 9% of the evaluations they found in their literature review having been conducted using "some form of traditional expert-based test" (p.17), how expertise was determined or defined was not clarified in the paper.

Discussion of the target audience included the gaming community – that is, other gamers (e.g. the Video Game Guide (Koehler et al., 2017) or Steam (Lin et al., 2019) community), 'consumers' or 'customers' (Eberhard et al., 2018; Ho & Tu, 2012) or purchasing decision-makers (Kasper et al., 2019), 'players' (Livingston et al., 2011) or 'users' (Wang & Goh, 2020). Other audiences included game developers (Caserman et al., 2020; Coleman & Talib, 2014; Kasper et al., 2019; Lin et al., 2019; Petri et al., 2016), academics/researchers (Calderón & Ruiz, 2015; Caserman et al., 2020; Faizan et al., 2019; Kirschner & Williams, 2014; Petri et al., 2016; Yáñez-Gómez et al., 2017) or instructors/educators (Faizan et al., 2019; Petri et al., 2016). Some reviews had multiple audiences. Petri et al. (2016) targeted both game developers and/or instructors; Kasper et al. (2019) both purchasers and/or game developers. Three articles – (Calderón & Ruiz, 2015), (Petri et al., 2016), and (Faizan et al., 2019) – provided reviews for researchers and/or educators/practitioners; and one targeted developers and/or researchers (Caserman et al., 2020).

*4.2 Why are EduXR/ Game reviews being provided?*

From analysis we found purposes for performing a game review vary greatly and depends on the stakeholder groups which the reviewing authors belong to. The major reason for reviewers to perform a game review is to inform other players of their personal game play experiences as well as their insights to the design of the game. The sharing of opinions was also dependent on the reviewers' background which influenced the purpose of performing the review.

Eight studies provided details where players who review a game choose to direct their work at peers with a common interest (Koehler et al., 2017). Players provided opinions which influenced the perception of the game and the play experience (Livingston et al., 2011), and to influence purchasing decisions (Eberhard et al., 2018; Kasper et al., 2019), to communicate with the designers and developers of the game (Lin et al., 2019), and to encourage the player to think as a designer to gain greater insights to understanding the game (Kirschner & Williams, 2014). One article (Petri & Gresse von Wangenheim, 2017) concludes that in sharing opinions on education games "there exists a need for the identification of more consistent and uniform patterns for systematically evaluating educational games in order to obtain valid results for communicating improvement to game quality." Yáñez-Gómez et al. (2017) shares that the "most prevalent objective when evaluating games is to measure the satisfaction, acceptance and engagement of users, thereby resulting in improvement of the learning effectiveness and therapeutic effect when users are motivated."

Expert reviewers also provided details on the quality of the game play in two articles (Coleman & Talib, 2014; Shiratuddin & Thabet, 2011) through quality control of "design aspects of specific criteria such as real-time rendering, real-time walkthrough, interactivity, multi-participatory feature, lighting and collision detection." (Shiratuddin & Thabet, 2011) concludes that contributing to a design review system by an expert "will reduce the time necessary to complete the review phases, thus, shorten the project life-cycle" in game development. 5 articles provide details about experts' reviewers making recommendations and providing critical assessments to game design (Coleman & Talib, 2014; Kasper et al., 2019; Koehler et al., 2017; Livingston et al., 2011; Yáñez-Gómez et al., 2017).

Educational reviewers for serious games share reviews which have additional goals and require reviews that assess how well they achieve these goals. Three articles (Calderón & Ruiz, 2015; Faizan et al., 2019; Shiratuddin & Thabet, 2011) describe that educators contribute reviews to "evaluate the usefulness of simulation games as well as their students' readiness, motivation, and learning outcomes" and to achieve specific goals by "giving educators an opportunity to identify students' perceptions of their change in knowledge, skills, attitudes and behavior" after playing a game. (Faizan et al., 2019) provided "strategies for educational simulation games by presenting suitable assessment types for the pre-game, in-game, and postgame assessment of simulation games" to ensure that educational content in a game is accurate (Coleman & Talib, 2014). The educator review contributes directly to assess the success of an educational game as suggested by (Petri et al., 2016), particularly if the structure of the review is specified in advance (Petri et al., 2016).

Game owners and developers provide game reviews to share information on their product (Petri et al., 2016). This includes sharing qualitative elements in the review to provide insight into game play, design, and game-based training principles that will enhance the ratings and commercial value of their property (Lin et al., 2019; Wang & Goh, 2020).

Researchers and academics evaluate games and provide reasoning about their properties (Faizan et al., 2019; Kirschner & Williams, 2014; Koehler et al., 2017; Wang & Goh, 2020). Where (Faizan et al., 2019) shared an evaluation strategy within different game phases with intent for use by other researchers to assess educational simulation games. Using the gameplay review method by (Kirschner & Williams, 2014), the study concludes that using empirical and interpretive data game play engagement can be measured giving deeper understanding of the game.

**Conclusion**

A scoping review of the scholarly literature provides a useful starting point to consider what is currently known about the nature of educational game reviews. The results highlighted key characteristics of who conducts eduXR reviews and why they do so. eduXR reviews can be described by the authors details, their expertise and their intended audience. Further, authors provide eduXR reviews for a variety of reasons, which includes: to share opinion, for quality control, to describe education aspects, to rate and measure an experience, and for game choosing what experience to play. The analysis did not capture the motivation of the reviewer, such as economic interest. Further, we are aware that our scoping review outcomes sourced articles from academic journals/ conferences, and those articles provided a focus on video game reviews, rather than explicitly eduXR. Our analysis is limited by data coming from this one perspective. What are the needs in eduXR that is different to serious games? EduXR experiences are expected to provide an educational opportunity, with the eduXR audience choosing an experience based on their purpose and need for educational outcomes. EduXR reviews are to provide this opportunity for the audience to understand educational elements, yet a valid review structure is required to enable audience agency in choice of eduXR experience. We expect from our scoping review, that there are anticipated similarities between review system expected by audiences of games and eduXR, with the outcomes of scoping review applying across these application spaces.

What is currently unknown via scoping review outcome is information about the actual audience of these eduXR reviews. Currently the audience is understood through reviewer intent, who the reviewer intends to target with the review. The audience, or who engages with the eduXR reviews is unknown, however Hendrickx (2022) lens of audience agency and engagement on three levels, micro, meso, and macro can assist in positioning eduXR reviews. Audience may be considered as an aggregate of users on the macro level, positioned as a commercial entity on the meso level or as an individual user on the mirco level. Within this framework agency includes both creating reviews but also using or responding to reviews, often as deliberate forms of action but sometimes incidental when on an individual level. In our scoping review, the literature identified four broad categories of intended audience, which consists of consumers/ gamers, game developers, academics/researchers, and educators/ practitioners/ instructors. When considering the audience of eduXR, we can consider what the literature defines in term of the audience of serious games. Calderón & Ruiz (2015) describes that serious games are often evaluated by adults studying in higher education. The use of serious games in the primary and secondary school context is considered in the literature, however present a smaller cohort of evaluative studies. In addition, these evaluations are often based on small (n = 40) groups of participants evaluating an experience. Amateur reviews predict game reputation, success, demand and sales (Santos et al., 2019). Santos et al. (2019) argued that online customer reviews are important for consumer choice. Individual or micro agency appears common in eduXR reviews. However, audience agency is not solely influenced by our own interests but purposefully directed by the platforms we visit and information we receive (Hendrickx 2022). Further Hendrickx (2022) argued that "there is currently a major disconnect between individuals' decision-making processes to engage with media content and similar processes on publishing or disseminating said content at the editors' end". Further exploration is required to understand who the audience of eduXR reviews is, and to understand how audience agency occurs with eduXR reviews. Importantly we need to understand how eduXR consider the needs of the learner/teacher audience, and how this audience can exhibit agency over the review content provided. Reviewers of eduXR have more authority than the consumer of review content and provide a journalist role when contributing an eduXR review. Similarly teachers/ instructors are more invested in educational value of the review than many consumers of media content, and in the accuracy of the content than many stakeholders in media content creation.

The scoping review outcomes indicate an opportunity to expand the review base, to consider what is included in reviews from a broader perspective. For example, review systems in our scoping study did not consider the use of eduXR in an educational context from a privacy and legal perspective. For educational institutions, particularly those supporting children, such information is essential when considering deploying an experience. In addition, how eduXR content changes over time was not considered, nor was how social media platforms form part of the eduXR review system. Overall, further insight is also required into what distinguishes eduXR from other applications and how educational outcomes are emphasized in the review process by all stakeholders.